\documentclass[%
 reprint,hidelinks,nofootinbib,
 amsmath,amssymb,
 aps,
]{revtex4-2}

\usepackage{graphicx}
\usepackage{dcolumn}
\usepackage{bm}
\usepackage{csquotes}
\usepackage{xcolor}
\usepackage{hyperref}
\usepackage{comment}

\begin{document}

\preprint{APS/123-QED}

\newcommand{\physrep}{{Phys. Rep.}}

\title{Axion sourcing in dense stellar matter via CP-violating couplings}

\author{Filippo Anzuini$^{1,2}$}\email{filippo.anzuini@gmail.com}
\author{Antonio G\'omez-Ba\~{n}\'on$^{3}$}
\author{Jos\'e A. Pons$^{3}$}
\author{Andrew Melatos$^{1,2}$}
\author{Paul D. Lasky$^{4, 5}$}

\affiliation{$^{1}$School of Physics, The University of Melbourne, Parkville, Victoria 3010, Australia
}
\affiliation{$^{2}$ Australian Research Council Centre of Excellence for Gravitational Wave Discovery (OzGrav), The University 
of Melbourne, Parkville, Victoria 3010, Australia}
\affiliation{$^{3}$ Departament de Física Aplicada, Universitat d'Alacant, 03690 Alicante, Spain}
\affiliation{$^{4}$School of Physics and Astronomy, Monash University, Clayton, Victoria 3800, Australia}
\affiliation{$^{5}$OzGrav: The ARC Centre of Excellence for Gravitational Wave Discovery, Clayton, Victoria 3800, Australia}

\date{\today}
\newcommand{\jcap}{{JCAP}}
\newcommand{\aap}{{A\&A}}
\newcommand{\mnras}{{MNRAS}}
\newcommand{\apjl}{{ApJL}}
\newcommand{\ssr}{{Space Science Reviews}}
\newcommand{\nphysa}{{Nucl. Phys. A}}

\begin{abstract}
Compact objects such as neutron stars and white dwarfs can source axion-like particles and QCD axions due to CP-violating axion-fermion couplings. The magnitude of the axion field depends on the stellar density and on the strength of the axion-fermion couplings. We show that even CP-violating couplings one order of magnitude smaller than existing constraints source extended axion field configurations. For axion-like particles, the axion energy is comparable to the magnetic energy in neutron stars with inferred magnetic fields of the order of $10^{13}$ G and exceeds by more than one order of magnitude the magnetic energy content of white dwarfs with inferred fields of the order of $10^{4}$ G. On the other hand, the energy stored in the QCD axion field is orders of magnitude lower due to the smallness of the predicted CP-violating couplings. It is shown that the sourced axion field can polarize the photons emitted from the stellar surface, and stimulate the production of photons with energies in the radio band. 
\end{abstract}

\maketitle

\section{Introduction}
Numerous experimental and theoretical research efforts are devoted to probing the phenomenology of axions, light pseudoscalar particles originally introduced to solve the strong CP problem \citep{Peccei_1977a, Peccei_1977b, Wilczek_1978, Weinberg_1978}. Alongside experiments in terrestrial laboratories \citep{Sikivie_1983, Sikivie_1984, Hagmann_1990, Smith_1999, SQUID_2010, CAST_2017, Du_2018, Knirck_2019, Kim_2019, Braine_2020, Lee_2020}, white dwarfs (WDs) and neutron stars (NSs) provide unique opportunities to test axion models in extreme environments, that cannot be replicated on Earth \citep{Raffelt_2008,Leinson_2014, Sedrakian_2016, Hamaguchi_2018, Sedrakian_2019, Buschmann_2021b}. 

Recently, it has been shown that NSs can \enquote{source} (i.e. produce) axions lighter than the standard QCD axion \citep{Hook_2018, Huang_2019, Zhang_2021, Di_Luzio_2021}, causing a large shift of the axion field expectation value in NS interiors. The same effect may occur for the QCD axion due to the appearance of \enquote{exotic} condensates at densities above the nuclear saturation density \citep{Balkin_2020}. The energy stored in the sourced axion field can be larger than the magnetic energy expected in magnetars and can trigger a magnetic dynamo mechanism that alters the standard magneto-thermal evolution of NSs \citep{Anzuini_2023}. However, large axion-field configurations in astrophysical bodies can be produced also via CP-violating axion-fermion interactions, which can be probed by laboratory experiments \citep{Irastorza_2018}. More in general, the effect of Beyond Standard Model light scalars on compact stars has been studied in \citep{Gao_2022}, where it is shown that in the case of a linear, strong Yukawa coupling with standard matter the sourcing of the light scalar can affect the mass-radius relation of stellar remnants. Also, \citep{BalkinWD_2022,Balkin_2023} have studied the impact of a sourced scalar on the stellar structure (or equation of state) of NSs and WDs.

In this work, we show that dense stellar objects such as NSs and WDs can source axion-like particles (ALPs) and QCD axions due to CP-violating couplings between axions and fermions, producing large axion-field configurations. For ALPs, we find that even CP-violating couplings one order of magnitude smaller than the current sensitivity limits of laboratory experiments \citep{Smith_1999,Geraci_2017, Irastorza_2018, Lee_2020, OHare_2020} shift the ALP field considerably. In NSs, the energy content stored in the axion field is comparable to the magnetic energy of NSs with inferred magnetic fields in the range $\lesssim 10^{13}$ G. In WDs, we show that the energy of the axion field is more than one order of magnitude larger than the magnetic energy of stars with fields of the order of $10^{4}$ G. In both NSs and WDs, the sourced ALP field can lead to potentially observable signatures. For the QCD axion, the smallness of the predicted CP-violating couplings reduces by orders of magnitude the energy content of the sourced axion field compared to the ALP case. 

This paper is structured as follows. In Section \ref{sec:axions_dense_matter} we review the equation of motion of axions at high densities, discuss the solution in different regimes, and study the energetics associated with the sourced axion field. The effects of axion sourcing on the photon emission of NSs are presented in Section \ref{sec:Implications}. The comparison with the axion field sourced via magnetospheric fields and additional possible observables are presented in the Discussion.

\section{Axions in dense matter}
\label{sec:axions_dense_matter}
We introduce the equation of motion for the axion field in Section \ref{eq:axion_EoM}, specializing in the case of ALPs in Section \ref{sec:ALP_EoM}, and their energetics in Section \ref{sec:energetics_ALPs}. We study the QCD axion in detail in Section \ref{sec:QCD_EoM}.

\subsection{Axion equation of motion}
\label{eq:axion_EoM}
We study axions in the high density regime, assuming that they interact only with neutrons, protons, and electrons.
The CP-conserving axion Lagrangian density reads \citep{Irastorza_2018, Di_Luzio_2020, Bertolini_2021}

\begin{align}
    \mathcal{L}^{CPC}_a = \ & \frac{1}{2}(\partial_\mu a)(\partial^\mu a) - \mathcal{V}(a) - \frac{g_{a\gamma}}{16\pi}a F_{\mu\nu}\tilde{F}^{\mu\nu} \nonumber \\ 
    & - a\sum_{j}g_{aj}(i\bar{\psi}_j\gamma^5\psi_j) \ .
    \label{eq:axion_lagrangian}
\end{align}
In Eq. \eqref{eq:axion_lagrangian}, $a$ is the axion field, $\mathcal{V}(a)$ denotes the axion potential, $F_{\mu\nu}$ is the electromagnetic strength tensor and $\tilde{F}_{\mu\nu}$ its dual. The index $j$ runs over the fermionic degrees of freedom interacting with axions (neutrons, protons, and electrons in our model, denoted by the fields $\psi_j$).

The quantity $ i\bar{\psi}_j\gamma^5\psi_j$ has contributions from the divergence of the spin density in the non-relativistic limit, which in turn depends on the
magnetic field strength, and to a lesser extent from the anomalous magnetic moment of the fermion (in case there is a non-vanishing electric field). Since the CP-conserving term couples the axion to the spin density of particles, one would need a high degree of polarization for the CP-conserving term to be comparable to the CP-violating term (discussed below). In the magnetar scenario, we expect $B \approx 10^{14}$ G and $T= 10^8 - 10^9$ K = $10-100$ keV, so that
$e \hbar B/m_nc \approx 1$ keV $\ll ~kT$. Thermal fluctuations wash out any significant spin-polarization of nucleons, while for electrons one has $e \hbar B/m_ec \approx$ few MeV. However, the CP-conserving term is several orders of magnitude smaller than the CP-violating term. Thus, in the following, we focus on axion sourcing only via the CP-violating axion-fermion interactions.

If there are sources of CP-violation in the Standard Model, axions acquire CP-violating couplings leading to the following Yukawa terms \citep{Irastorza_2018, Di_Luzio_2020, Bertolini_2021}
\begin{equation}
    \mathcal{L}^{CPV}_a = - a\sum_{j = n,p}\bar{g}_{aj}(\bar{\psi}_j\psi_j) \ ,
\end{equation}
where $\bar{g}_{aj}$ are the CP-violating coupling constants of axions with nucleons. Such couplings have been recently considered in the context of compact stars in \citep{Gao_2022}, which studied the effect on the mass-radius relation of NSs in the case of a light scalar linearly coupled to standard matter. Note that the QCD axion interacts only derivatively with electrons (i.e. only with CP-conserving interactions). We assume that ALPs participate in the same interactions as QCD axions to compare the two cases directly. 

We can now derive the Klein-Gordon (KG) equation for the axion field, reading \citep{Irastorza_2018}

\begin{align}
    \Box a + m^2_a a & =  \ \frac{g_{a\gamma}}{4\pi}\mathbf{E}\cdot\mathbf{B} - \sum_{j = n, p}\bar{g}_{aj}\langle\bar{\psi}_j\psi_j\rangle ,
    \label{eq:KG}
\end{align}
where $m_a$ denotes the axion mass, and $\mathbf{E}$, $\mathbf{B}$ denote respectively the electric and magnetic fields (we neglect the terms proportional to $g_{aj}$). Since axions are expected to be light ($m_a \lesssim 0.1$ eV), their typical Compton wavelength is macroscopic, and they interact with the mean-field values in dense matter of the field bilinears (i.e. $\langle\bar{\psi}_j\psi_j\rangle$). 

In the following, we use the notation $n_S = \sum_{j}\langle\bar{\psi}_j\psi_j\rangle$ for the scalar density in the KG equation. This can be obtained from \citep{Serot_1997}

\begin{equation}
    n_S = \frac{\gamma}{(2\pi)^3}\int_0^{k_F}{d^3k\frac{M_N^*}{E_N^*(k)}} ,
    \label{eq:scalar_density}
\end{equation}
where $M^*_{N}$ denotes the nucleon effective mass in dense matter, and $E_N^*(k) = \sqrt{k^2 + M^{*2}_{N}}$. The prefactor $\gamma$ is the spin-isospin degeneracy factor \citep{Serot_1997}, and $k_F$ denotes the Fermi momentum. 

For NSs, we adopt the GM1A equation of state (EoS) \citep{Gusakov_2014}, matched with the SLy4 equation of state in the crust \citep{Douchin_2001}. We focus on a NS model with mass $M_{NS} = 1.49 \ M_{\odot}$ and with radius $R_{NS} = 13.8$ km, containing only nucleons and leptons. For sufficiently low densities however (typically in the NS crust), the ground state of dense matter is given by nuclei, and we approximate the scalar density with its non-relativistic limit, i.e. the baryon density, similarly to the case of axion fields sourced by test masses in laboratory experiments (e.g. \citep{Arvanitaki_2014}). 
For WDs, we use a relativistic polytropic equation of state. We adopt a WD model with mass $M_{WD} = 0.7 \ M_{\odot}$ and $R_{WD} = 7.98 \times 10^3$ km. In this case, $n_S$ in Eq. \eqref{eq:scalar_density} is approximated with the baryon density.

\begin{figure*}
\includegraphics[width=5.8cm, height = 5.5cm]{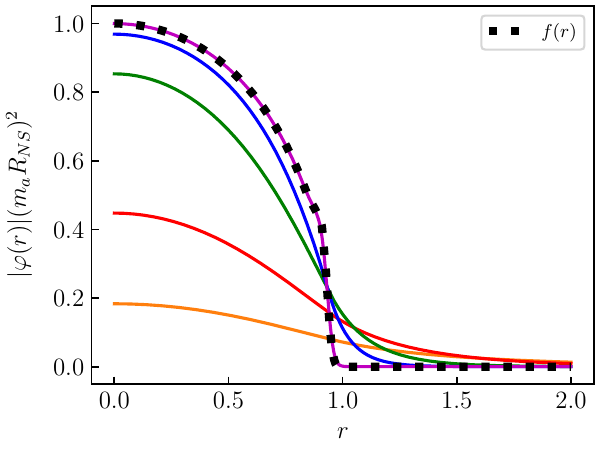}
\includegraphics[width=5.8cm, height = 5.5cm]{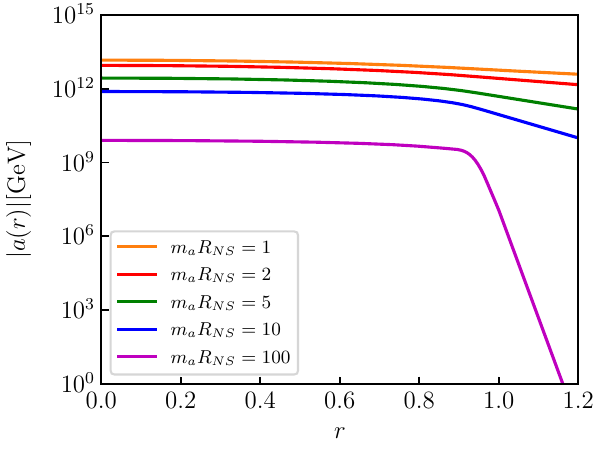}
\includegraphics[width=5.8cm, height = 5.5cm]{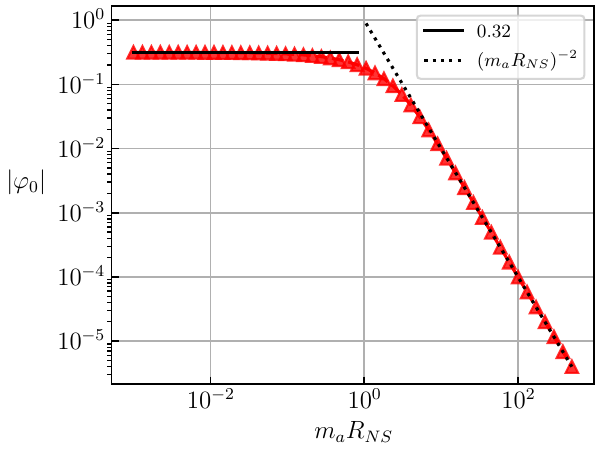}
\caption{\label{fig:ALP_NS_fields}  ALP field in NSs versus for different values of $m_a R_{NS}$. Left panel: radial profiles of $(m_a R_{NS})^2 |\varphi(r)|$ versus normalized radius for $m_a R_{NS} = 1, 2,5,10,100$ (see the legend in the central panel for the color labels). The dotted curve represents the $f(r)$ profile of the $M_{NS} = 1.49 \ M_{\odot}$ model, to which the solution converges as $m_a R_{NS}$ increases. Central panel: amplitude of the axion field $a$ in physical units (GeV) versus normalized radius. Right panel: scaling of the central value $\varphi_0$ with $m_a R_{NS}$.}
\end{figure*}


\subsection{Sourcing ALPs}
\label{sec:ALP_EoM}
For ALPs, the axion mass $m_a$ and the couplings $\bar{g}_{aj}$ can be treated as free parameters. Due to the high electrical conductivity of both NSs and WDs, the electric field is weak, with $E \approx (v/c) B$ (where the drift velocity of the electrons is $1 \ \textrm{km} \ \textrm{Myr}^{-1} \lesssim v \lesssim 10^3 \ \textrm{km}\ \textrm{Myr}^{-1}$ in the crust). In such conditions, the electromagnetic source term ($\propto \mathbf{E}\cdot\mathbf{B}$) in the KG equation is subdominant, and we neglect it. Note however that the electromagnetic term can give relevant contributions in the magnetosphere of both NSs and WDs \citep{Garbrecht_2018}. A full solution of the KG equation including the electromagnetic term requires accurate modeling of magnetospheric fields and is left for future work. We compare however our results obtained without the electromagnetic source terms with the ones found in \citep{Garbrecht_2018} in the following sections.

To discuss in detail how the solutions of the axion field scale with the relevant parameters, it is convenient to introduce the energy scale\footnote{The following results are valid for both NSs and WDs, and we use the generic notation $M, R$ to denote the mass and radius of a compact star. When considering specific NS and WD models, we write $M_{NS}, R_{NS}$ and $M_{WD}, R_{WD}$ respectively.}

\begin{equation}
e_a = \bar{g}_{aN}n_{S}(r = 0)R^2 \ ,
\label{eq:ea_scale}
\end{equation}
where $R$ is the star's radius and the scalar density is evaluated at the center of the star ($r = 0$). We seek solutions of the stationary KG equation in terms of the dimensionless variable $\varphi=a/e_a$. Expressing the radial coordinate in units of $R$ (that is, hereafter $r \equiv r/R$), the KG equation can be written as

\begin{equation}\label{eq:kgs}
\frac{\partial^2\varphi}{\partial r^2}+\frac{2}{r}\frac{\partial\varphi}{\partial r} - (Rm_{a})^2 \varphi = f(r)~,
\end{equation}
where $f(r)$ is the source term divided by $e_a$, which by definition decreases monotonically from $f(r=0)=1$.

We solve Eq. \eqref{eq:kgs} with the appropriate boundary conditions both, numerically and exactly. The exact formal solution is \footnote{Another useful way to write the solution is in terms of the Green's function: $$\varphi(r)=-\frac{1}{2m_aRr}\int_{0}^{1}dr'r'f(r')\left(e^{-m_aR|r'-r|}-e^{-m_aR|r'+ r|}\right)$$}

\begin{align}
\varphi(r)  = & \frac{1}{m_a R r} \int_{r}^{1}dr'r'f(r') \sinh[m_a R(r'-r)]   \nonumber \\
& -\frac{A}{r}e^{-m_a Rr} \ ,
\label{eq:gensolkg}
\end{align}
with
\begin{equation}
A= \frac{1}{m_a R} \int_{0}^{1}dr'r'f(r')\sinh(m_a Rr').
\end{equation}

Since the source function $f(r)$ vanishes for $r\ge 1$, the solution outside the star reduces to 
\begin{equation}
\varphi(r)= - \frac{A}{r}e^{-m_a Rr}~ \quad \mathrm{for} ~{r \ge 1}.
\end{equation} The constant $A$ can be evaluated for any given density profile and sets the surface value of the axion field. We can also evaluate the limit to obtain the value at the origin $\varphi_0$:
\footnote{
Since $f(r')$ is a decreasing, positive defined function with $f(0)=1$, we have
$|\varphi_0|  \leq \int_{0}^{\infty} x e^{-x}dx = 1$}
\begin{equation}
\varphi_0 = - \int_{0}^{1}dr' f(r')r'e^{-m_a Rr'} \ .
\label{theta0}
\end{equation}
One can recover the original dimensions for the axion field by multiplying $\varphi_0$ 
by the energy scale $e_a$.

It is interesting to discuss how the amplitude of the field behaves in the different axion-mass limits.
In the interior of the star, we find
\begin{eqnarray}
a(r) \approx \left\{ 
\begin{array}{cc}
  -\frac{e_a}{(m_a R)^2} f(r) & \mathrm{if} ~m_a R \gg 1 \\
  - \beta(r) e_a & \mathrm{if} ~m_a R \lesssim 1
\end{array}
\right.
\label{asymptotic}
\end{eqnarray}
where $\beta(r) \lesssim \frac{1}{2}$ is a function of order unity; for the particular NS model employed in this work we have $\beta(r=0)=0.32$. 
Note that for $m_a R \gg 1$ the maximum amplitude of the field scales as $(m_a R)^{-2}$. 

Here we take $\bar{g}_{aN} = 10^{-23}$, which is one order of magnitude smaller than the lower bounds reported in \citep{Irastorza_2018, Di_Luzio_2020} (see also references therein and the limits reported in \cite{AxionLimits}) for light ALPs. In NSs, this results in $e_a \approx 7.9 \times 10^{13}~ \mathrm{GeV}$. 

In Figure \ref{fig:ALP_NS_fields} we consider the NS case and show results of the stationary KG equation varying the parameter $m_a R_{NS}$ in the range $1 \leq m_a R_{NS} \leq 100$, corresponding to $ 10^{-11} \lesssim m_a/\textrm{eV}\lesssim 10^{-9}$. In the left panel of Figure \ref{fig:ALP_NS_fields}, we compare the radial profiles of the dimensionless axion field $(m_a R_{NS})^2 |\varphi (r)|$. The dotted curve represents the normalized scalar density profile of the NS model adopted in this work, which is almost indistinguishable from the axion field profile for $m_a R_{NS} \gtrsim 20$. The amplitude of the field in physical units (GeV) is plotted in the central panel, which shows that the field decays exponentially outside the star. 
We also note that further decreasing the dimensionless axion mass $m_a R_{NS}$ does not lead to an arbitrarily large axion field. Instead, it converges to a solution that tries to minimize gradients (which become the dominant contribution to the total energy in the regime $m_aR_{NS}\ll 1$) inside the star. In the intermediate regime, $m_a R_{NS} \lesssim 1$, the solution of Eq.\eqref{eq:kgs} finds an optimal balance between gradient and potential terms. In the opposite limit ($m_a R_{NS} \gg 1 $), the gradient terms can be neglected and the axion field follows the relation reported in Eq. \eqref{asymptotic}. 
In the right panel, we show the scaling of the central value ($\varphi_0$) with $m_a R_{NS}$ and find that $\varphi_0 \approx 0.32$ for $m_a R_{NS} \lesssim 1$ and $\varphi_0 \propto (m_a R_{NS})^{-2}$ for large values.
We show a wide range of $m_a R_{NS}$ for completeness, although values of $m_a$ in the range 
$ 10^{-13} \ \textrm{eV} \lesssim m_a \lesssim 10^{-11}$ eV (that is, $ 10^{-2} \lesssim m_a R_{NS} \lesssim 1$) are, in principle, excluded by black hole superradiance arguments \citep{Arvanitaki_2015, Cardoso_2018}.

\subsection{Energetic considerations for ALPs}
\label{sec:energetics_ALPs}
We now determine the energy stored in the ALP field sourced by NSs. The energy density reads $\mathcal{H} = (m_a a)^2/2 + (\nabla a)^2/2$. From the behavior of the solutions described in Eq. (\ref{asymptotic}), we find
\begin{eqnarray}
\mathcal{H} \approx \left\{ 
\begin{array}{cc}
  \frac{1}{2} (m_a R_{NS})^{-2} \left( \frac{e_a}{R_{NS}} \right)^2 & \mathrm{if} ~m_a R_{NS} \gg 1 \\
  \frac{1}{2} \beta^2 (m_aR_{NS})^2\left( \frac{e_a}{R_{NS}} \right)^2 & \ \mathrm{if} ~m_a R_{NS} \lesssim 1~
\end{array}
\right.
\label{asymptotic2}
\end{eqnarray}
at $r = 0$. Note that the maximum value of the energy density is smaller than $\left( e_a/R_{NS}\right)^2$, which is also of the same order as the interaction energy 
$\mathcal{H}_{int} = \bar{g}_{aN}n_{S}(r = 0) a \approx  e_a a/R_{NS}^2$. 
Using Eq. (\ref{eq:ea_scale}), we have 
\begin{equation}
\left( \frac{e_a}{R_{NS}} \right)^2 \approx  2\times10^{-7} ~\mathrm{MeV~fm}^{-3} 
\approx 3 \times 10^{26} ~\mathrm{erg~cm}^{-3} \ ,
\label{eaunits}
\end{equation}
which is comparable to the magnetic energy density of NSs (given by $B^2/8\pi$) with inferred fields in the range $B \in [10^{12}, 10^{13}]$ G for $m_aR_{NS} \lesssim 1$. For $m_aR_{NS} \gg 1$, the axion energy density scales as $(m_a R_{NS})^{-2}$.

\begin{figure*}
\includegraphics[width=9cm, height = 7cm]{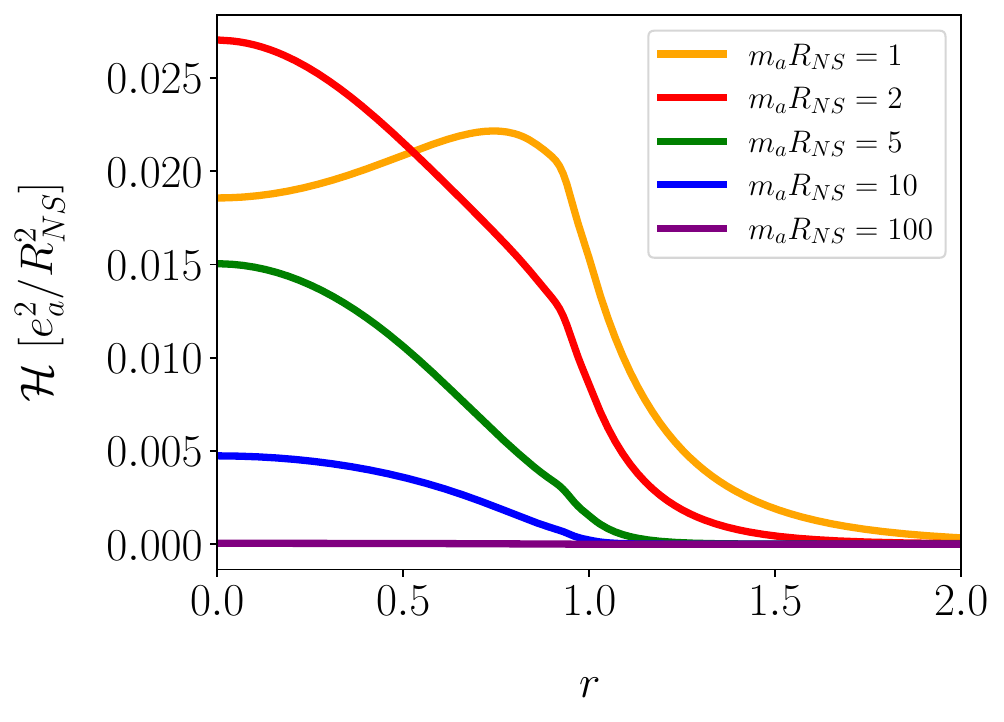}%
\includegraphics[width=9.cm, height = 7.cm]{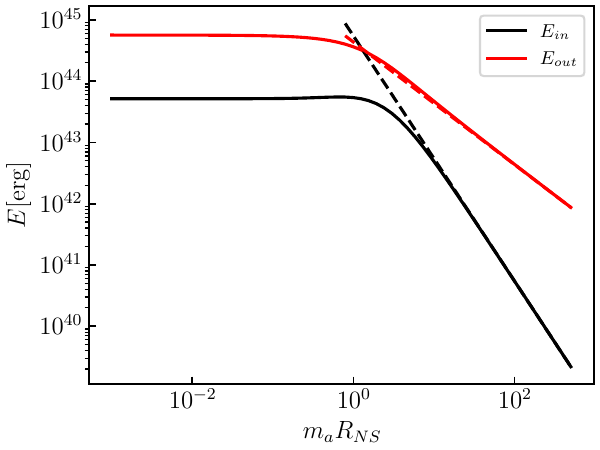}
\caption{Energy stored in the ALP field. Left: energy density profiles of the axion field for different values of $m_a R_{NS}$, in units of $e_a^2/R_{NS}^2$ (see Eq. (\ref{eaunits})). Right:
total energy integrated over the star volume inside ($E_{in}$) and outside ($E_{out}$) the star. The maximum energy stored in the axion field ($E_{in} \approx 7 \times 10^{43}$ erg) is reached for $m_a R_{NS} \lesssim 1$. The dashed lines show the scalings $(m_a R_{NS})^{-1}$ and $(m_a R_{NS})^{-2}$ in the limit $m_a R_{NS} \gg 1$.
\label{fig_energy}
}
\end{figure*}

The left panel of Figure \ref{fig_energy} compares different radial profiles of the energy density for different choices of the parameter $m_a R_{NS}$. The typical energy density stored in the axion field is of order $10^{-2} \left(e_a/R_{NS} \right)^2 \approx 10^{24
} ~\mathrm{erg~cm}^{-3}$ for $m_a R_{NS}$ of order unity, and scales as $(m_a R_{NS})^{-2}$ for heavier masses, as discussed above. The right panel of Figure \ref{fig_energy} shows the total energy integrated over the star volume

\begin{equation}
E_{in} = 4 \pi R_{NS}^3 \int_0^{1} dr r^2 \mathcal{H}(r)
\end{equation}
which is a quantity of the order of
$10^{-1} e_a^2 R_{NS} \approx 7\times 10^{43}$ erg for the NS model considered here and $m_a R_{NS} = 1$, which is comparable to the total magnetic energy in NSs with poloidal-dipolar fields and inferred fields of the order $B \approx 10^{13}$ G \citep{Potekhin_2020}. For completeness, we also include in the figure the total energy of the axion field outside the star
\begin{equation}
E_{out} = 4 \pi R_{NS}^3 \int_1^{\infty} dr r^2 \mathcal{H}(r)  ~.
\end{equation}
We note that the maximum energy available outside the star is of the order of $10^{44}$ erg, which can be relevant for magnetospheric physics but still irrelevant for long-range gravitational effects (compared to $M_\odot  c^2$).

One can obtain a rough estimate of the total axion energy for the axion field sourced by WDs as follows. Since in WDs one has $m_a R_{WD} \gg 1$, 
we obtain
\begin{eqnarray}
\mathcal{H} \approx 
  \frac{1}{2} \left(\frac{ \bar{g}_{aN}n_{S}}{m_a} \right)^2
  \approx 10^{10} \mathrm{~erg~ cm}^{-3} \ ,
\end{eqnarray} 
where we used $m_a=10^{-11}$ eV and $n_s = 10^{30}$ cm$^{-3}$.
Therefore, the energy of the axion field in WDs is $E_{in, WD} \approx 10^{37} \mathrm{~erg}$.
While the axion energy is considerably lower than the NS case, it is still comparable to the magnetic energy expected in some white dwarfs. The 
magnetic fields of WDs are in the range $10^{3} \lesssim B/\textrm{G} \lesssim 10^{9}$, with approximately $90\%$ of the WD population having fields below $10^{6}$ G \citep{Wickramasinghe_2005, Ferrario_2015}. This leads to an estimated magnetic energy of $10^{38} \left( B / 10^6{\mathrm G}\right)^2$ erg assuming for simplicity a poloidal-dipolar magnetic field.

\subsection{QCD axion}
\label{sec:QCD_EoM}
We now turn to the QCD axion. We calculate finite density corrections to the axion mass following \citep{Hook_2018, Balkin_2020, Di_Luzio_2021}. The QCD axion potential in vacuo reads \citep{Cortona_2016}

\begin{equation}
\mathcal{V}_0(a) = -m_\pi^2 f_\pi^2 \left[\sqrt{1 - \frac{4m_um_d}{(m_u+m_d)^2}\sin^2\left(\frac{a}{2f_a}\right)} - 1\right]\ ,
\label{eq:axion_potential}
\end{equation}
(we subtract a constant so that the potential vanishes when $a = 0$), where $m_{\pi} = 135$ MeV and $f_{\pi} = 93$ MeV denote the neutral pion mass and pion decay constant respectively, $m_u$ and $m_d$ are the up and down quark masses in vacuo respectively and $f_a$ is the axion decay constant.
The QCD axion potential in vacuo $\mathcal{V}_0$ in Eq. \eqref{eq:axion_potential} is proportional to the vacuum quark condensate $\langle \bar{q}q\rangle_0$ (where $q$ denotes the quark field) via the Gell-Mann-Oakes-Renner relation

\begin{equation}
    \langle \bar{q}q \rangle_0(m_u + m_d) = - m^2_{\pi}f^2_{\pi} \ .
\end{equation}
In dense matter, the expectation value of the quark condensate $\langle\bar{q}q\rangle_n$ changes with respect to the in-vacuo condensate. Using the Hellmann-Feynman theorem \citep{Cohen_1992}, one gets

\begin{equation}
    2m_q\bigg(\langle\bar{q}q\rangle_n - \langle\bar{q}q\rangle_0\bigg) = m_q\frac{d \mathcal{E}}{d m_q} \ ,
\end{equation}
where $m_q$ denotes the average, bare mass of up and down quarks and $\mathcal{E}$ is the energy density of the system, which depends on the EoS. For the GM1A EoS and the NS model studied in this work, one obtains

\begin{align}
    \frac{\langle\bar{q}q\rangle_n}{\langle\bar{q}q\rangle_0} = & \ 1 - \frac{\sigma_N}{m^2_\pi f^2_\pi}\bigg[\frac{m^2_\sigma}{g^2_{{\sigma N}}}(M_N - M^*_N) \nonumber \\
    &+ \frac{g_3}{g^3_{\sigma N}}(M_N - M^*_N)^2  + \frac{g_4}{g^4_{\sigma N}}(M_N - M^*_N)^3  \nonumber \\
    & + \chi_\sigma\frac{m_\sigma}{g^2_{{\sigma N}}}(M_N - M^*_N)^2 -\chi_{\omega}\frac{g^2_{{\omega N}}n^2_N}{m^3_\omega}  \nonumber \\
    &-\chi_{\rho}\frac{g^2_{{\rho N}}(n_p-n_n)^2}{4 m^3_\rho}\bigg].
    \label{eqn:condensate_medium}
\end{align}
In Eq. \eqref{eqn:condensate_medium}, we take $\sigma_N = 59$ MeV \citep{Cohen_1992, Balkin_2020}, and use $M_N$ to denote the bare mass of nucleons. $g_{\sigma N}$ and $m_\sigma$ are the coupling constant of the $\sigma$ meson with nucleons and the mass of the $\sigma$ meson respectively, and $g_3$ and $g_4$ denote the coupling constants for cubic and quartic self-interactions of the $\sigma$ meson respectively. The last two lines in Eq. \eqref{eqn:condensate_medium} include the contributions of the vector mesons $\omega$ and $\rho$, with masses $m_\omega$ and $m_\rho$ and couplings $g_{\omega N}$ and $g_{\rho N}$ to nucleons respectively. For the quantities $\chi_l$ (with $l = \sigma, \omega, \rho$), we follow \citep{Cohen_1992} and set $\chi_l \approx m_l/M_N$. $n_p$, $n_n$, and $n_N$ represent the proton, neutron, and nucleon number densities respectively (in particular, $n_N$ coincides with the baryonic density in the absence of hyperons). For simplicity, in the following we use the notation $\langle\bar{q}q\rangle_n/\langle\bar{q}q\rangle_0 = 1-F_n(r)$.

Eq. \eqref{eqn:condensate_medium} can be used to determine how finite density corrections change the QCD axion potential $\mathcal{V}(a)$ in dense matter, i.e.

\begin{equation}
    \mathcal{V}(a) \approx \big[1 - F_n(r)\big] \mathcal{V}_0(a) \ .
    \label{eq:QCD_potential_dense_matter}
\end{equation}
The effective axion mass is then 

\begin{equation}
    m^*_a = m_{\pi} f_\pi\frac{\sqrt{z}}{f_a(1+z)}\sqrt{1 - F_n(r)} \ ,
    \label{eqn:effective_mass}
\end{equation}
with $z = m_u/m_d = 0.48$. Given $m^*_a$, the radial profile of the QCD axion field in dense matter is determined using Eq. \eqref{eq:QCD_potential_dense_matter} to solve the KG equation.

We now calculate the magnitude of the QCD axion field in NSs. The CP-violating couplings for the QCD axion are expected to be in the range \citep{OHare_2020}

\begin{equation}
    10^{-29}\bigg(\frac{10^9 \ \textrm{GeV}}{f_a}\bigg) \lesssim \bar{g}_{aN} \lesssim 10^{-21}\bigg(\frac{10^9 \ \textrm{GeV}}{f_a}\bigg) \ .
    \label{eq:CP_coupling_range}
\end{equation}
The resulting coupling strength $\bar{g}_{aN}$ is weaker than in the ALP case studied above, the sourced axion field is in general smaller, and the corresponding phenomenology is harder to test (as discussed in Section \ref{sec:discussion}). 

The natural energy scale is now $f_a$. We define $\phi = a/f_a$, and the KG equation becomes:

\begin{align}
\frac{\partial^2\phi}{\partial r^2}+\frac{2}{r}\frac{\partial\phi}{\partial r}   = \left(\frac{e_s}{f_a}\right)^2\bigg[& f(r) + \left(\frac{m_a^*Rf_a}{e_s}\right)^2 \nonumber \\
&  \times \frac{\sin \phi}{\sqrt{1-\frac{4z}{(1+z)^2}\sin^2(\phi/2)}}\bigg],
\label{eq:kgs2}
\end{align}
with $\phi = a/f_a$ and $e_s = \sqrt{e_af_a}$.

In the following, we take $\bar{g}_{aN} = 10^{-12} ~\mathrm{GeV}/f_a$ for the QCD axion, i.e. the upper bound in Eq. \eqref{eq:CP_coupling_range}. With our choice of the CP-violating couplings, the energy scale $e_s$ is fixed to $e_s \approx 2.8\times10^{12}~\mathrm{GeV}$.

We also note that the coefficient of the last term on the right-hand side of Eq. (\ref{eq:kgs2}) is independent of the scales $R$ and $f_a$, i.e. 
$$
\left(\frac{m_a^*Rf_a}{e_s}\right)^2 \approx 
2\times10^{10}\left[1-F_n(r)\right] \ .
$$
\begin{figure*}
\includegraphics[width=9.5cm, height = 7.5cm]{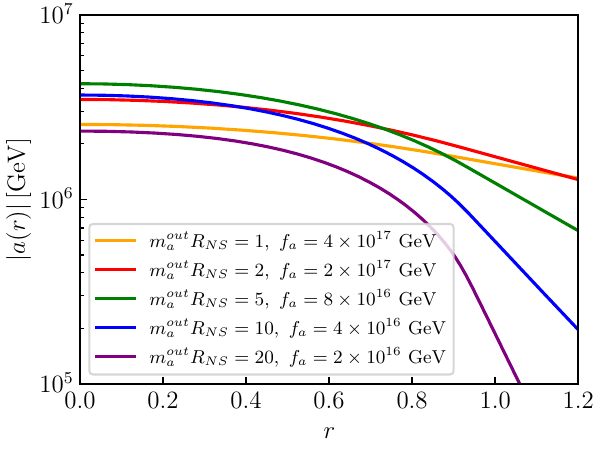}%
\includegraphics[width=9.5cm, height = 7.5cm]{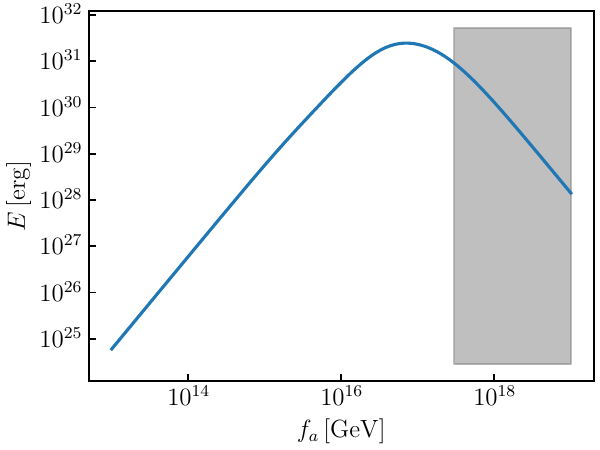}
\caption{QCD axion field sourced by NSs and corresponding energy. Left: QCD axion field sourced by NSs for different values of $f_a$. In the legend, $m^{out}_a$ denotes the QCD axion mass in vacuo. Right:
total energy integrated over the star volume versus $f_a$. The gray shaded region corresponds to the $f_a$ interval excluded by stellar black-hole superradiance arguments \citep{Arvanitaki_2015, Marsh_2017}.
\label{fig:QCD_NS_fields}
}
\end{figure*}

For the EoS adopted in this work, $(m_a^*)^2$ remains positive ($1-F_n >0$) and the solution can be well approximated expanding the QCD axion potential in Eq. \eqref{eq:axion_potential} and using only the $(a/f_a)^2$ contribution. Then, as with ALPs, we can consider the limiting case $m_a^*R \gg 1$. For $m_a^*R \gg 1$, since the solution to the KG equation satisfies $\phi \ll 1$ ($a\ll f_a$), it can be accurately approximated by 
$$
a = -\frac{e_a}{(m_a^* R)^2} f(r) 
$$
(note that now $m_a^*$ is not constant but a function that varies with $r$).

The case with $1-F_n =0$ at some critical radius $r_c$ in the star interior has been studied in the literature \citep{Balkin_2020} (see \citep{Hook_2018, Huang_2019, Zhang_2021} for a non-standard QCD axion model). In such conditions, the axion condenses, and the solution to the KG equation must also be obtained numerically. However, in the limit $m_a^*R \gg 1$, the solution approaches a step function, where a condensate of amplitude $a = \pi f_a$ in the core ($r<r_c$) drops exponentially for $r>r_c$.

We now report the field profiles and total energy stored in the axion field for the QCD axion model. In Figure \ref{fig:QCD_NS_fields} we study the QCD axion sourced by NSs for representative values of $f_a$. In the left panel we vary $f_a$ in the range $f_a \in [2\times 10^{16}, 4 \times 10^{17}]$ GeV. We also report the dimensionless quantity $m^{out}_aR_{NS}$, where $m^{out}_a$ denotes the mass of the QCD axion in vacuo. Contrarily to the ALP case, varying $f_a$ affects not only the axion mass but also the couplings (and hence the magnitude of the source term in the KG equation). As a consequence, we find that the QCD axion field at the center of the star is largest for $f_a = 8 \times 10^{16}$ GeV. For higher values of $f_a$, the axion-fermion couplings ($ \propto f^{-1}_a$) decrease, reducing the source term in the KG equation and the magnitude of the axion field. Compared to the ALP field calculated in Section \ref{sec:ALP_EoM}, the QCD axion field is approximately six orders of magnitude smaller. In the right panel of Figure \ref{fig:QCD_NS_fields} we calculate the total energy stored in the axion field for several values of $f_a$. Our results show that the total energy is orders of magnitude smaller than the ALP case, attaining approximately $10^{31}$ erg for $f_a = 8 \times 10^{16}$ GeV.

In general, the smaller the axion field and its gradients, the lower the energy stored in the axion field. This limits the possible effects of the axion field on typical observables of compact stars, as discussed below.

\section{Implications for electromagnetic emission}
\label{sec:Implications}

\begin{figure*}
\includegraphics[width=11.cm, height = 8.cm]{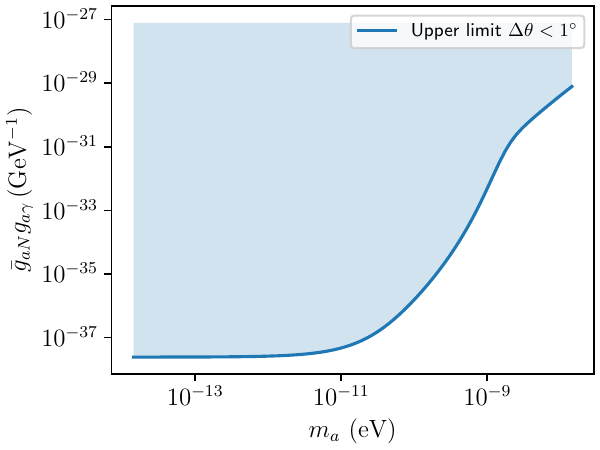}%
\caption{Bounds on the product $\bar{g}_{aN}g_{a\gamma}$ obtained from the condition $\Delta\theta<1^{\circ}$. The blue shaded region is excluded since it implies $\Delta\theta > 1^{\circ}$, which is larger than the sensitivity of polarimeters searching for light polarization from NSs \cite{Moran_2013, Liu_2020}.}
\label{fig_birefringence}
\end{figure*}
The sourcing of axions via CP-violating interactions in dense matter leads to a wealth of possible observational implications. These range from the polarization of light propagating through the axion-dense medium, to the emission of photons via axion decay, as well as to modifications of the standard magneto-thermal evolution due to axion contributions in Maxwell's equations \citep{Sikivie_1983, Wilczek_1987, Visinelli_2013, Kim_2019}. In this section, we study the polarization of light emitted from the stellar surface and the production of photons via axion-photon interactions.

We first focus on the polarization of photons emitted from NSs, along the lines of \cite{Poddar_2020}. The circular polarization modes of photons propagating through the axion-dense magnetosphere have dispersion relations that differ by an axion-induced correction (birefringence). Following \citep{Poddar_2020}, one can use the axion-modified Maxwell's equations to obtain the birefringent angle $\Delta\theta$, which reads

\begin{equation}
\Delta\theta = \frac{1}{2}\int^{\infty}_{R_\gamma} dr \left(k^+_r - k^-_r\right) \ ,
\end{equation}
where $R_\gamma \approx R_{NS}$ is the photospheric radius, $k^{\pm}_r = \omega \mp g_{a\gamma}(\nabla_r a)/2$ is the radial momentum of the photon and $\omega$ is the photon energy (we neglect the metric factors for simplicity). For the exponentially decaying profiles outside the star, one can write

\begin{equation}\label{eq:biref_angle}
    \Delta\theta = \frac{g_{a\gamma}}{2}e_a|\varphi(1)|=\frac{g_{a\gamma}\bar{g}_{aN}n_{s}(r=0)R^2_{NS}}{2}|\varphi(1)| ~ \ .
\end{equation}

Eq. \eqref{eq:biref_angle} can be used to put constraints on the product $\bar{g}_{aN}g_{a\gamma}$, similarly to the constraints obtained in \citep{Gau_2023} on the product of $g_{a\gamma}$ and the CP-conserving axion-nucleon coupling $g_{aN}$ using magnetar polarization data. For example, imposing $\Delta\theta < 1^{\circ}$ (i.e. that the birefringent angle due to axions is smaller than the best expected sensitivity of polarimeters \cite{Moran_2013, Liu_2020}) and using the results in Section \ref{sec:ALP_EoM} one obtains

\begin{align}
   g_{a\gamma}\bar{g}_{aN} &< \frac{\pi}{90}\frac{1}{\left(n_{s}(r=0)|\varphi(1)|R^2_{NS}\right)}~,
   \label{eq:biranglebound}
\end{align}
where
\begin{equation}
    |\varphi(1)|= \frac{e^{-m_aR_{NS}}}{m_a R_{NS}} \int_{0}^{1}dr'r'f(r')\sinh(m_a R_{NS}r') ~. 
    \label{eq:bound2}
\end{equation}
We display the corresponding bound in Figure \ref{fig_birefringence},
where the blue shaded region would be excluded in the hypothetical case that observations of a number of pulsars allow us to place the limit $\Delta\theta < 1^{\circ}$.

The polarization of light can also be induced by the magnetic field in the magnetosphere. The difference with the axion-induced polarization is that the latter affects systematically all photons passing through the axion-dense magnetospheric region. In contrast, the polarization due to the magnetic field is wavelength-dependent \cite{Wang_2011, Poddar_2020}. 

\begin{figure*}
\includegraphics[width=11.cm, height = 8.cm]{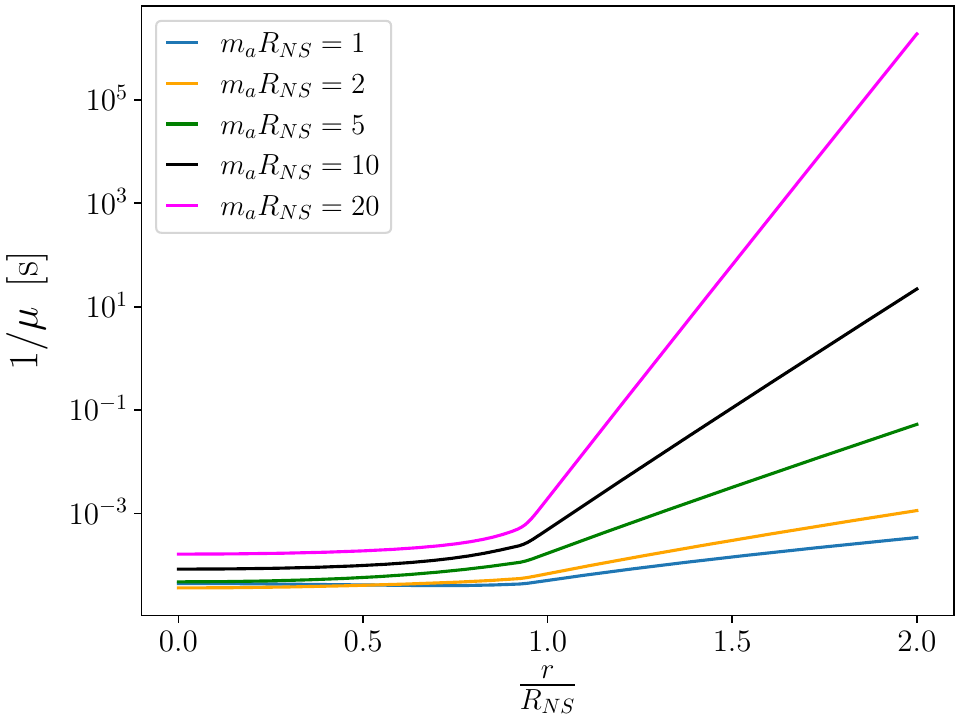}%
\caption{Time-scale for the exponential growth of the number density of photons produced via axion-photon interaction for different values of the dimensionless axion mass $m_a R_{NS}$. The photons are produced with typical momenta $k = \mathcal{O}(m_a)$, which for the axion masses considered in our work leads to the production of photons with energies in the radio band. The axion-photon coupling is fixed to $g_{a\gamma} = 10^{-13}$ GeV$^{-1}$.
\label{fig_mu}
}
\end{figure*}

We now turn to the axion decay emission processes. It has been found that the decay rate to two photons is enhanced in axion- and photon-dense environments \citep{Caputo_2019, Carenza_2020} (\enquote{stimulated emission}). We apply such results to our case and show that axions sourced in NSs can potentially result in an excess of emission in the radio band. 
As discussed in \citep{Caputo_2019, Carenza_2020}, axions may decay into two photons with typical momenta\footnote{This is a standard result in cosmology, where an oscillating axion field coupled with gauge fields produces dark photon abundances which grow exponentially in time (cf. \citep{Agrawal_2018}).} $k = m_a/2$ with a width in momentum space given by $\Delta k \approx g_{a\gamma} a m_a$. Note that $\Delta k \ll k$ and the photon radiation is assumed to be essentially monochromatic. Although the typical decay timescale is larger than the age of the Universe, the corrections due to stimulated emission result in an enhanced effective decay rate of axions into two photons given by \citep{Carenza_2020}

\begin{equation}
\Gamma_{eff} = \frac{g^2_{a\gamma} m^3_a}{64\pi}\left(1 + \frac{8\pi^2a n_\gamma}{g_{a\gamma}m_a^2 n_a}\right) \ ,
\label{eq:eff_rate}
\end{equation}
where $n_a$ and $n_\gamma$ are the axion and photon number densities respectively. In Eq. \eqref{eq:eff_rate}, the second term in the brackets is the enhancement factor, which depends on the photon number density (for $n_\gamma = 0$ the decay rate coincides with the spontaneous decay rate).

We should note that the equation above, as outlined in \citep{Carenza_2020}, was initially derived for homogeneous axion condensates. However, it has also been applied in the same reference to inhomogeneous axion clumps, in a scenario where $m_a R_* \gg 1$, with the axion clump's radius $R_*$ of the order of $10^{5}$ km. In principle, the results derived for an homogeneous background can be applied to the inhomogeneous case if the 
wavelength of the particles considered is shorter than the lengthscale of variation of the background, say $\frac{1}{m_a} \ll R_{NS}$.

From the Boltzmann equation, it can be shown that the photon number density evolves according to 

\begin{equation}
\dot{n}_\gamma = 2 \Gamma_{eff}n_a \ ,
\label{ndot}
\end{equation}
which, in the absence of other reactions, leads to an epoch with an exponentially growing photon number density, i.e.

\begin{equation}
n_\gamma(t) = n_{\gamma}(0)e^{\mu t} \ ,
\end{equation}
with

\begin{equation}\label{eq:mu_timescale}
\mu = \frac{\pi}{4}g_{a\gamma}m_a \delta a \approx \frac{\pi}{4}g_{a\gamma}\sqrt{2\mathcal{H}} \ .
\end{equation}
In Eq. \eqref{eq:mu_timescale}, $\delta a$ represents the amplitude of the time-dependent perturbation of a background field. 

The factor $m_a \delta a$ arises from the time derivative of the axion field \citep{Hertzberg:2018zte}, assuming $\dot{a} \approx m_a \delta a$ (oscillation frequencies of order $m_a$).
The amplitude of the perturbations depends on the particular mechanism considered (thermal fluctuations, p-modes or sound waves modifying the source in the KG equation, etc.). 
Considering time-dependent oscillations of the order of the static background amplitude, we can approximate $\mathcal{H}$ with the energy density studied in Section \ref{sec:axions_dense_matter}. Thus, this value of $\mu$ must be taken as an upper limit, and a more precise calculation would require to solve numerically the Klein-Gordon coupled with the hydrodynamics or magnetohydrodynamics system of equations with a given perturbation.

In Figure \ref{fig_mu}, we plot the timescale $\mu^{-1}$ versus the normalized radius of the star and for different values of the dimensionless parameter $m_a R_{NS}$ (we fix the axion-photon coupling to $g_{a\gamma} = 10^{-13}$ GeV$^{-1}$). 
We show results for $m_a R_{NS} \geq 1$, i.e. for the case in which the scale of spatial variation is larger than the wavelength of the axions, similarly to the case studied in \citep{Carenza_2020}. The timescale is far shorter in the interior of the star compared to the exterior region. However, the photons produced in the opaque interior will be quickly reabsorbed due to interaction with the NS constituents. 
We note that this timescale is proportional to $(\delta a)^{-1}$, so the reader can rescale the plot according to their expectations. For example, perturbations of the order of one-thousandth of the background field will lead to timescales of $10^{-2}-1$ s in the region close to the star surface.

On the contrary, photons produced in the transparent region surrounding the star could potentially contribute to some observable signal with durations from ms to s, typical of pulsar phenomenology. 
In this case, to account for the flux losses, Eq. (\ref{ndot}) must be modified with a negative term proportional to $n_{\gamma}$. This second characteristic time is equal to the light crossing time of the region of interest \citep{Kephart1995} and the exponential growth could be suppressed.
For $m_a R_{NS} \approx 1$, the produced photons have energies of the order of $10^{-11}$ eV, i.e. in the long wavelength radio band ($<0.1$ MHz). Unfortunately, radiation in this band is absorbed by the Earth's atmosphere so no information is available from ground-based radio telescopes. Note, however, that in the range $m_a R_{NS} > 1000$ the expected emission lies in the GHz region and, although the energy stored in ALPs decreases with $m_a$, it is still consistent (see Figure \ref{fig_energy}, right panel) with the typical energetics of the intriguing and popular Fast Radio Bursts (FRBs),  $10^{40}-10^{42}$ erg \cite{Zhang2018}, and much larger than other transient radio-burst phenomena in pulsars. In summary, axion-decay is enhanced via stimulated emission in the inner magnetosphere, which opens an interesting possibility that deserves a more thorough study.

\section{Discussion}
\label{sec:discussion}

In compact stars, the ALP field is shifted by several orders of magnitude compared to its vacuum value, even for CP-violating couplings one order of magnitude smaller than the ones that can be probed by current axion direct searches \citep{Irastorza_2018, AxionLimits} (cf. also \citep{Gao_2022, BalkinWD_2022, Balkin_2023} for the effects of light scalars strongly coupled with dense matter on stellar remnants). In such conditions, several observational signatures may become detectable. These can be used to constrain the axion parameter space, e.g. the axion mass and couplings.

We find that CP-violating interactions between axions and fermions source large-scale axion field configurations with energy density $\mathcal{H} \approx 10^{24}$ erg cm$^{-3}$ for $ 1 < m_a R_{NS} \lesssim 10$ outside of a NSs (cf. Figure \ref{fig_energy}). This is several orders of magnitude larger than the typical energy density associated with axions sourced by magnetospheric fields for an axion-photon coupling $g_{a\gamma} = 10^{-13}$ GeV$^{-1}$ \citep{Garbrecht_2018}.  We note that the sourcing of axions due to magnetospheric fields produces axion configurations that extend for radii of the order of $20$ km \citep{Garbrecht_2018}. For axions sourced by CP-violating couplings, the axion field falls exponentially outside the star. However, for $m_aR_{NS} < 10^{-2}$ (corresponding to $m_a < 10^{-13}$ eV), the axion field outside of the star decreases exponentially on a length-scale $1/m_a \approx 10^{3}$ km, leading to an axion-dense medium that encompasses the NS magnetosphere, and whose energy density is orders of magnitude larger than the one obtained in \citep{Garbrecht_2018}. 

The dense axion field configurations found in this work can lead to a wealth of observational features. For example, photons moving through an axion-dense environment can be polarized due to axion-photon interactions \citep{Tkachev_2015, Garbrecht_2018, Poddar_2020}. This may affect both the light emitted from the stellar surface, as the compact object cools down, and the light emitted by a second source transiting in the surroundings of the compact object. For the axion parameters used in this work, the birefringence angle is smaller than $1^{\circ}$ (the typical sensitivity of polarimeters) for $g_{a\gamma} \lesssim 10^{-15}$ GeV$^{-1}$ and for $m_a R_{NS} \gtrsim 1$. Moreover, since the dense axion medium sourced by the star is threaded by magnetospheric fields, axion-photon conversions may lead to some form of radio burst from NSs on short timescales $\ll 1$ s. For our particular scenario of ALPs, the energy and timescales observed in FRBs would be consistent with $m_a R_{NS} \approx 1000$. 

The coupling of axions to photons can lead to further consequences for compact objects. Maxwell's equations are modified by axion corrections \citep{Sikivie_1983, Wilczek_1987, Kim_2019} which can alter the standard magnetic evolution of compact stars \citep{Anzuini_2023}. For example, axions may form secondary electromagnetic fields and currents, and the latter may be dissipated providing the star with an additional source of internal heat besides the standard Joule heating in the absence of axions \citep{Pons_2007, Vigano_2013, Pons_2019, Anzuini_2021, Anzuini_2022}. Due to the comparable axion and magnetic energy densities in standard NSs with inferred fields of the order $10^{13}$ G, it is difficult to assess whether axions can leave an imprint on the magnetic evolution of standard NSs. On the other hand, the ALP energy exceeds by roughly two or three orders of magnitude the magnetic energy of WDs with field strengths of the order of $10^{4}$ G (which constitute the majority of the WD population \citep{Wickramasinghe_2005, Ferrario_2015}), and axion-induced corrections may alter the magnetic field configuration more easily (compared to the NS case).

The phenomenology related to the QCD axion is more difficult to test, given the lower energies compared to the ALP case. The energy density associated with the QCD axions outside NSs is below the typical energy density of axion configurations sourced by the magnetospheric fields (of the order of $9.5 \times 10^{-2}$ g cm$^{-3}$, as found in \citep{Garbrecht_2018}, at least for $g_{a\gamma} = 10^{-13}$ GeV$^{-1}$), making it hard to constrain the parameter space of QCD axions sourced by CP-violating couplings and the related phenomenology.

The present work can be extended by including the electromagnetic backreaction on the axion field, which can become relevant when the dimensionless axion mass is in the range $m_a R_{NS} \gtrsim 100$ and the axion field sourced by CP-violating couplings drops by several orders of magnitude outside the star. In general, detailed models of the magneto-thermal evolution of NSs and WDs are necessary to calculate the axion profile including the effect of magnetospheric fields. This requires a self-consistent numerical investigation of the interior and exterior magnetic field, which must be coupled to the thermal evolution of compact stars including the relevant cooling channels (e.g. neutrino emissivity in NSs \citep{Yakovlev_1999, Potekhin_2015, Raduta_2018, Raduta_2019, Anzuini_2021, Anzuini_2022}).

\begin{acknowledgments}
FA thanks Anthony W. Thomas, Koichi Hamaguchi, Federico Bianchini and Davide Guerra for fruitful discussions. We also thank the anonymous referee for useful comments. FA, PDL and AM are supported by the Australian Research Council (ARC) Centre of Excellence for Gravitational Wave Discovery (OzGrav), through project number CE170100004. PDL is supported through ARC Discovery Project DP220101610. JAP and AG acknowledge support from the Generalitat Valenciana grants ASFAE/2022/026 (with funding from NextGenerationEU PRTR-C17.I1) and CIPROM/2022/13, 
and from the AEI grant PID2021-127495NB-I00 funded by MCIN/AEI/10.13039/501100011033 and by “ESF Investing in your future”.
\end{acknowledgments}

\bibliography{apssamp}

\end{document}